\documentclass[showkeys,showpacs,nofootinbib,superscriptaddress]{revtex4-2}
\usepackage{amsmath}
\usepackage{amssymb}
\usepackage{graphicx}
\usepackage{float}
\usepackage{xcolor}
\usepackage{physics}
\usepackage{ulem}

\allowdisplaybreaks

\begin{document}

\title{
Color field configuration between three static quarks
}

\author{
Vladimir Dzhunushaliev
}
\email{v.dzhunushaliev@gmail.com}
\affiliation{
Department of Theoretical and Nuclear Physics,  Al-Farabi Kazakh National University, Almaty 050040, Kazakhstan
}
\affiliation{
Institute of Experimental and Theoretical Physics,  Al-Farabi Kazakh National University, Almaty 050040, Kazakhstan
}
\affiliation{
Academician J.~Jeenbaev Institute of Physics of the NAS of the Kyrgyz Republic, 265 a, Chui Street, Bishkek 720071, Kyrgyzstan
}

\author{Vladimir Folomeev}
\email{vfolomeev@mail.ru}
\affiliation{
Institute of Experimental and Theoretical Physics,  Al-Farabi Kazakh National University, Almaty 050040, Kazakhstan
}
\affiliation{
Academician J.~Jeenbaev Institute of Physics of the NAS of the Kyrgyz Republic, 265 a, Chui Street, Bishkek 720071, Kyrgyzstan
}


\begin{abstract}
Within Yang-Mills-Proca theory with external sources in the form of three static quarks, regular, finite energy solutions are obtained. It is shown that color electric/magnetic fields have two components: the first part is a gradient/curl component, respectively, and the second part is a nonlinear component. It is shown that the color electric field has a Y-like spatial distribution provided by   three static quarks. Such a Y-like behavior arises because the gradient component of the electric field is present. The nonlinear component of the electric field is a curl one, and it appears because the vector potential sourced by a solenoidal current is present. The color magnetic field is purely curl one, since its nonzero color components do not contain a nonlinear component; this results in the fact that its force lines lie on the surface of a torus. It is shown that the results obtained are in satisfactory agreement with the results obtained in lattice calculations in quantum chromodynamics. To discuss such an agreement, we have shown that the Yang-Mills-Proca equation can be obtained from the Lagrangian describing a gluon condensate varying in space. Also, we compare the energy profile obtained by us with that obtained in lattice calculations with a static potential. 
\end{abstract}


\date{\today}

\maketitle

\section{Introduction}

At the present time lattice calculations in nonperturbative quantum chromodynamics (QCD) have achieved a great success (see, e.g., the reviews  \cite{Constantinou:2020hdm,Cichy:2018mum,Ding:2015ona} and references therein). Considerable results have been obtained: (a)~a comparison of  perturbative global PDF analyses and the lattice QCD calculations was carried out; (b)~QCD calculations on the thermodynamics of strong-interaction matter were performed; (c)~the QCD phase transition was investigated; (d)~an equation of state was proposed, etc. 

In Ref.~\cite{Alexandrou:2017oeh}, within lattice QCD, the nucleon spin carried by valence and sea quarks and gluons is determined. The papers~\cite{DiGiacomo:1990hc,Bali:1994de} study the distribution of a field between two quarks and show  that a color electric field created by these quarks is confined in a flux tube. Refs.~\cite{Bornyakov:2004uv,Koma:2017hcm,Sakumichi:2015rfa,Takahashi:2000te,Takahashi:2002bw,Suganuma:2023mml} study the distribution of quantum SU(3) gauge fields created by three static quarks. 

In lattice calculations (see, e.g., Ref.~\cite{Bornyakov:2004uv}), it is shown that, in QCD, there are well-ordered distributions of color electric and magnetic fields creating $\Delta$- and Y-like distributions of the color electric field. This suggests that there can exist differential equations describing such distributions. 
Searching for such equations is, of course, a quite complicated problem. One may assume that such equations must follow from the quantum Yang-Mills equations in the process of nonperturbative quantization as some approximation for such  quantization. We assume here that the Yang-Mills-Proca equations could be such equations, and discuss this in Sec.~\ref{dis_con}.

At the present time Proca theories (massive Yang-Mills theories) are under quite active  investigation. For example,  Refs.~\cite{Herdeiro:2017fhv,Dzhunushaliev:2019kiy,Dzhunushaliev:2019uft,Dzhunushaliev:2021vwn} 
study gravitating Proca fields, for which regular solutions are obtained. In Refs.~\cite{DeFelice:2016yws,Nakamura:2017dnf} the cosmological implications of generalized Proca theories are investigated. The paper~\cite{Minamitsuji:2016ydr} studies static and spherically symmetric solutions in the class of generalized Proca theories with non-minimal coupling to the Einstein tensor.

\begin{figure}[H]
	\centering
	\includegraphics[width=0.4\linewidth]{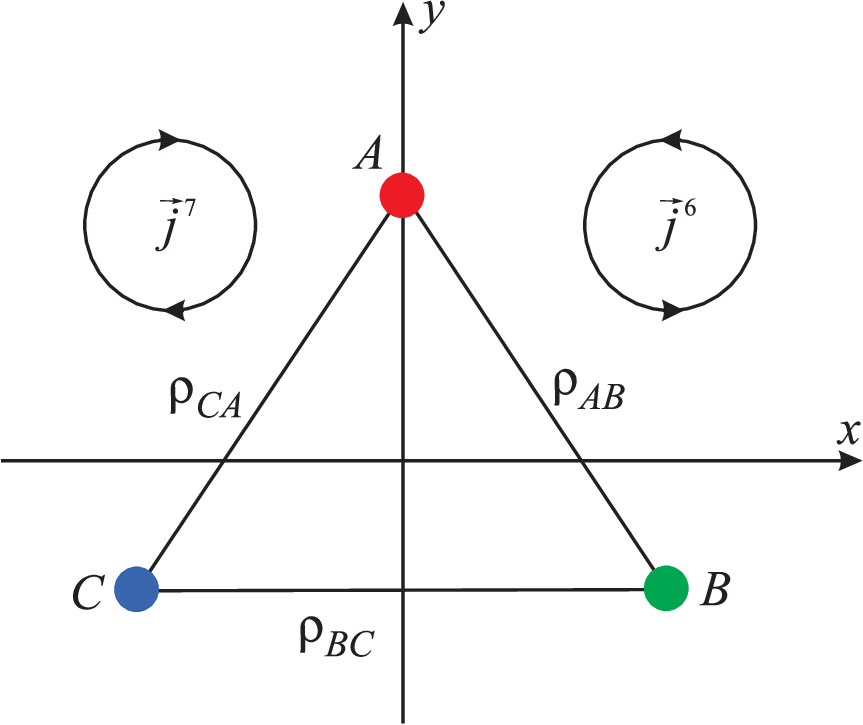}
	\caption{A sketch of distribution of static quarks and charge/current densities creating color electric, $\vec{E}^a$, and magnetic, $\vec{H}^a$, fields.}
\label{triangleABC}
\end{figure}

Our purpose here is to investigate a distribution of SU(3) non-Abelian field within Yang-Mills-Proca theory with external sources (see Fig.~\ref{triangleABC}): (a)~three static quarks located at the vertices  of an equilateral triangle $ABC$ which create a Y-like distribution of a color electric field; (b)~color charges distributed along the sides of this triangle which are needed for the existence of a nonlinear component of such electric field; 
(c)~toroidal currents (shown in Fig.~\ref{triangleABC}) which are necessary for the appearance of a color vector potential whose presence, in turn, results in the appearance of a curl component of the electric field.

We wish to show that Yang-Mills-Proca theory enables one to obtain satisfactory agreement with the results obtained in lattice calculations in QCD. This means that such theory may be regarded as some approximation for the procedure of nonperturbative quantization in QCD. In such case, there arises an interesting question of what physical approximations must be done in this procedure to derive Yang-Mills-Proca theory. To clarify this question, we will show that the Yang-Mills-Proca equation can be obtained  from the Lagrangian describing a SU(3) gluon condensate varying in space 
and discuss what approximations must be done to obtain necessary nonperturbative quantum corrections leading to the required result. 

\section{Starting Lagrangian and equations}

The Lagrangian describing a system consisting of a non-Abelian SU(3) Proca field $A^a_\mu$
sourced by color charges and currents can be taken in the form (hereafter, we work in natural units $c=\hbar=1$)
\begin{equation}
	\mathcal L =  - \frac{1}{4} F^a_{\mu \nu} F^{a \mu \nu} +
	\frac{m^2}{2}	A^a_\mu A^{a \mu} - A^a_\mu j^{a \mu} .
	\label{0_10}
\end{equation}
Here
$
F^a_{\mu \nu} = \partial_\mu A^a_\nu - \partial_\nu A^a_\mu +
g f_{a b c} A^b_\mu A^c_\nu
$ is the field strength tensor for the Proca field of mass $m$, where $f_{a b c}$ are the SU(3) structure constants; $g$ is the coupling constant; $ j^{a \mu}=\left\{j^{a t}, -\vec{j}^{\,a}\right\}$ are color four-currents;
$a,b,c = 1,2, \dots, 8$ are color indices; $\mu, \nu = 0, 1, 2, 3$ are spacetime indices. 

Using \eqref{0_10}, the corresponding field equations can be written in the form
\begin{equation}
	D_\nu F^{a \mu \nu} - 
	m^2 A^{a \mu} \equiv 
	\frac{1}{\sqrt{-\mathcal G}} \frac{\partial}{\partial x^\nu}\left(\sqrt{-\mathcal G}F^{a \mu \nu}\right) 
	+ g f_{a b c} A^b_\nu F^{c \mu \nu}
	- m^2 A^{a \mu} = - j^{a \mu},
\label{0_20}
\end{equation}
where $\mathcal G$ is the determinant of the spacetime metric
and the energy density is
\begin{equation}
	\begin{split}
		\varepsilon = &\frac{1}{2} \left( E^a_i \right)^2 +
		\frac{1}{2} \left( H^a_i \right)^2 +
		m^2\left( 
		A^{a 0} A^a_0 -
		\frac{1}{2}  A^a_\alpha A^{a \alpha}
		\right) ,
	\end{split}
\label{energy_dens}
\end{equation}
where $i=1,2,3$ and $E^a_i$ and $H^a_i$ are the components of the color electric and magnetic field strengths, respectively.

To find the SU(3) non-Abelian field in Yang-Mills-Proca theory, we employ the following ansatz in Cartesian coordinates: 
\begin{align}
\begin{split}
	A^a_\mu = & \frac{1}{g} \left\lbrace 
			h^a, v^a,u^a, w^a 
		\right\rbrace , \quad a= 3,6,7,8 , 
\\
	A^n_\mu = & 0, \quad n = 1, 2, 4, 5 , 
\end{split}
\label{gauge_pot}
\end{align}
where $h^{a}, v^a, u^a,$ and $w^a$ are functions of $x, y, z$ and $A^a_\mu$ belongs to the subgroup of SU(3).

For these potentials, we have the following color electric, $\vec{E}^a$, and magnetic, $\vec{H}^a$, fields:
\begin{align}
	E^a_i = & - \frac{1}{g} \left( \grad{h^a}\right)_i + g f_{abc} A^b_t A^c_i , 
\label{el_fields}\\
	H^a_i = & \left( \curl{\vec{A}^a} \right)_i + g f_{abc} \epsilon_{i j k} A^b_j A^c_k 
	= \left( \curl{\vec{A}^a} \right)_i + g f_{abc} \left[ \vec{A}^b \times \vec{A}^c \right]_i , 
\label{magn_fields}
\end{align}
where $i, j, k = 1, 2, 3$. These expressions imply that the color electric and magnetic fields have linear and nonlinear components. The color electric field $E^a_i$ has a gradient component, $-\left( \grad{h^a}\right)$, 
and the magnetic field $H^a_i$ has a curl component, $\left( \curl{\vec{A}^a} \right)$, just as in ordinary electrodynamics; also, there are nonlinear components for the electric field, $g f_{abc} A^b_t A^c_i$, and for the magnetic field, $g f_{abc} \vec{A}^b \times \vec{A}^c$. 
 
Calculating the covariant derivative of the left- and right-hand sides of Eq.~\eqref{0_20}, we have 
\begin{equation}
-m^2 \,\div{\vec{A}^a} = -\div{\vec{j}^a} + g f_{abc} A^b_\mu j^{c \mu} ,
\label{divergence}
\end{equation}
where $\div{\vec{A}^a}\equiv\partial_i A^a_i$ and $\div{\vec{j}^a}\equiv\partial_i j^{a}_i$. To eliminate mixed derivatives in the equations for the magnetic fields, we find partial derivatives of $\vec{A}^a$ from Eq.~\eqref{divergence} and substitute them into these equations. Note that for the self-consistent problem with fundamental (spinor and scalar) fields the divergence of the right-hand side is identically equal to zero.

\section{Y-like distribution of the color electric field for three static quarks}
\label{Y_distribution}

\subsection{Preliminary comments
}

\begin{figure}[H]
\centering
\includegraphics[width=0.3\linewidth]{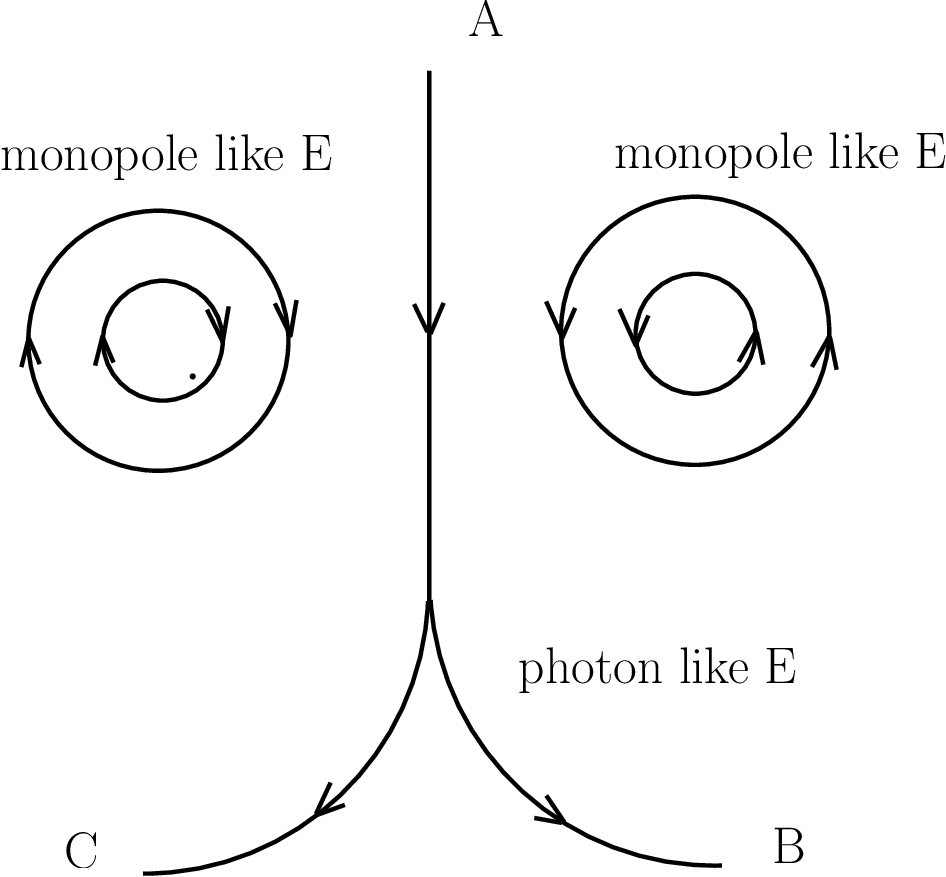}
\caption{A sketch of the Y-like distribution of the color electric field obtained in lattice calculations in Ref.~\cite{Bornyakov:2004uv}.}
\label{Y_E}
\end{figure}

In lattice calculations \cite{Bornyakov:2004uv}, it is shown that in QCD with three static quarks  located at the vertices of the triangle $ABC$ the color electric field has photon- and monopole-like components. The first one (photon) has a Coulomb-like behavior provided by the quarks located at the vertices of the triangle $ABC$. The monopole component is a curl one and, as pointed out in lattice calculations, it is created by a solenoidal, toroidal-like  current which flows along the tube and crosses the  plane of the figure perpendicularly, as sketched in Fig.~\ref{Y_E}. 

Our purpose here is to show that a similar distribution of force lines of the color electric field can be obtained in non-Abelian Yang-Mills-Proca theory with sources: 
\begin{itemize}
\item Located at the vertices of the triangle $ABC$ (three static quarks) and creating a Y-like profile of the gradient component of the color electric field:  
\begin{equation}
	\vec{E}^{3,8}_{\text{grad}} = - \frac{1}{g} \nabla{h^{3,8}}. 
\label{E_8_grd}
\end{equation}
This component corresponds to the first term in the expression for the color electric field given by Eq.~\eqref{el_fields}. 
\item Distributed along the sides of the triangle $ABC$ (see Fig.~\ref{triangleABC}). These charge densities $\rho^{6, 7}$ are needed for the creation of the components $A^{6, 7}_t$  of the potential. According to the expression~\eqref{el_fields}, the color electric field has two components: the gradient component $- \left( \grad{h^a}\right)$, which is called the photon part in Ref.~\cite{Bornyakov:2004uv}, and the nonlinear component $g f_{abc} A^b_t A^c_i$, which is called the monopole component in Ref.~\cite{Bornyakov:2004uv}. For  $\vec{E}^{3,8}$, the nonlinear component is represented by the expression
\begin{equation}
	\vec{E}^{3,8}_{\text{nonlinear}} = \frac{\alpha}{2 }  \left( h^6 \vec{A}^7 - h^7 \vec{A}^6\right)  , 
\label{E_8_nl}
\end{equation}
where $h^{6, 7} = g A^{6, 7}_t$; $\alpha = -1$ for $a=3$ and $\alpha = \sqrt{3}$ for $a=8$. 
\item Which are toroidal-like  currents $\vec{j}\,^{6, 7}$ (see Fig.~\ref{triangleABC}) needed for the appearance of the color vector potentials $\vec{A}^{6, 7}$ appearing in the expression~\eqref{E_8_nl}. 
\end{itemize}

The corresponding equations~\eqref{eqn_10}-\eqref{eqn_60} describing a spatial distribution of the fields are  given in Appendix~\ref{full_eqns}, since they are rather cumbersome. 

\subsection{Numerical method and boundary conditions}

To solve numerically the set of sixteen partial differential equations \eqref{eqn_10}-\eqref{eqn_60}, we rewrite
 them in terms of the dimensionless variables
$$
\left(x,y,z\right)\to \frac{1}{m}\left(x,y,z\right), \quad \left(h^a, v^a, u^a, w^a\right)\to m \left(h^a, v^a, u^a, w^a\right),
\quad  j^{a \mu} \to  \frac{m^3}{g}j^{a \mu} 
$$
and introduce compactified coordinates
\begin{equation}
\label{comp_coord}
	\bar{x} = \frac{2}{\pi} \arctan\left(\frac{x}{c_k}\right),  
	\bar{y} = \frac{2}{\pi} \arctan\left(\frac{z}{c_k}\right),
	\bar{z} = \frac{2}{\pi} \arctan\left(\frac{z}{c_k}\right),  
\end{equation}
which map the infinite region $(-\infty;\infty)$ onto the finite interval $[-1; 1]$. Here $c_k$ is an arbitrary constant which is used to adjust the contraction of the grid. In our calculations, we typically take $c_k=4$.

Technically, Eqs.~\eqref{eqn_10}-\eqref{eqn_60} are discretized on a grid consisting of $35 \times 35 \times 20$ grid points covering  the integration region $-1\leq \bar x, \bar y \leq 1$ and $0\leq \bar z \leq 1$
[given by the compactified  coordinates~\eqref{comp_coord}]. The resulting set of nonlinear algebraic equations is then solved by using a modified Newton method. The underlying linear system is solved 
with the Intel MKL PARDISO sparse direct solver \cite{pardiso} and the CESDSOL library\footnote{Complex Equations-Simple Domain partial differential equations SOLver, a C++ package developed by I.~Perapechka,
see Refs.~\cite{Herdeiro:2019mbz,Herdeiro:2021jgc}.}.

Taking into account that the problem under consideration is symmetric with respect to the  $z = 0$ plane, one can conclude that profiles of the fields must be either symmetric or antisymmetric with respect to the $z = 0$ plane. Then the boundary conditions can be taken in the form
\begin{align}
	\left. \pdv{h^{3,6,7,8}}{z}\right|_{z = 0} = & \left. \pdv{v^{6,7}}{z}\right|_{z = 0} = 
	\left. \pdv{u^{6,7}}{z}\right|_{z = 0} = 0, 
\nonumber\\
	\left. v^{3,8}\right|_{z=0} = & \left. u^{3,8}\right|_{z=0} = \left. w^{3,6,7,8}\right|_{z=0} = 0 .\nonumber
\end{align}
In turn, asymptotically (as $(x,y,z)\to \pm \infty$), all the above functions tend to 0. For such boundary conditions, the $\mathbb Z_2$ symmetry with respect to the  $z = 0$ plane results in the fact that the $z$-components of the electric and magnetic fields are
$E^{3,6,7,8}_z(x,y,z=0) = H^{3,6,7}_z(x,y,z=0) = 0$. 

\subsection{Numerical results and discussion}
\label{num_res}

\begin{figure}[t]
\centering
\includegraphics[width=1.\linewidth]{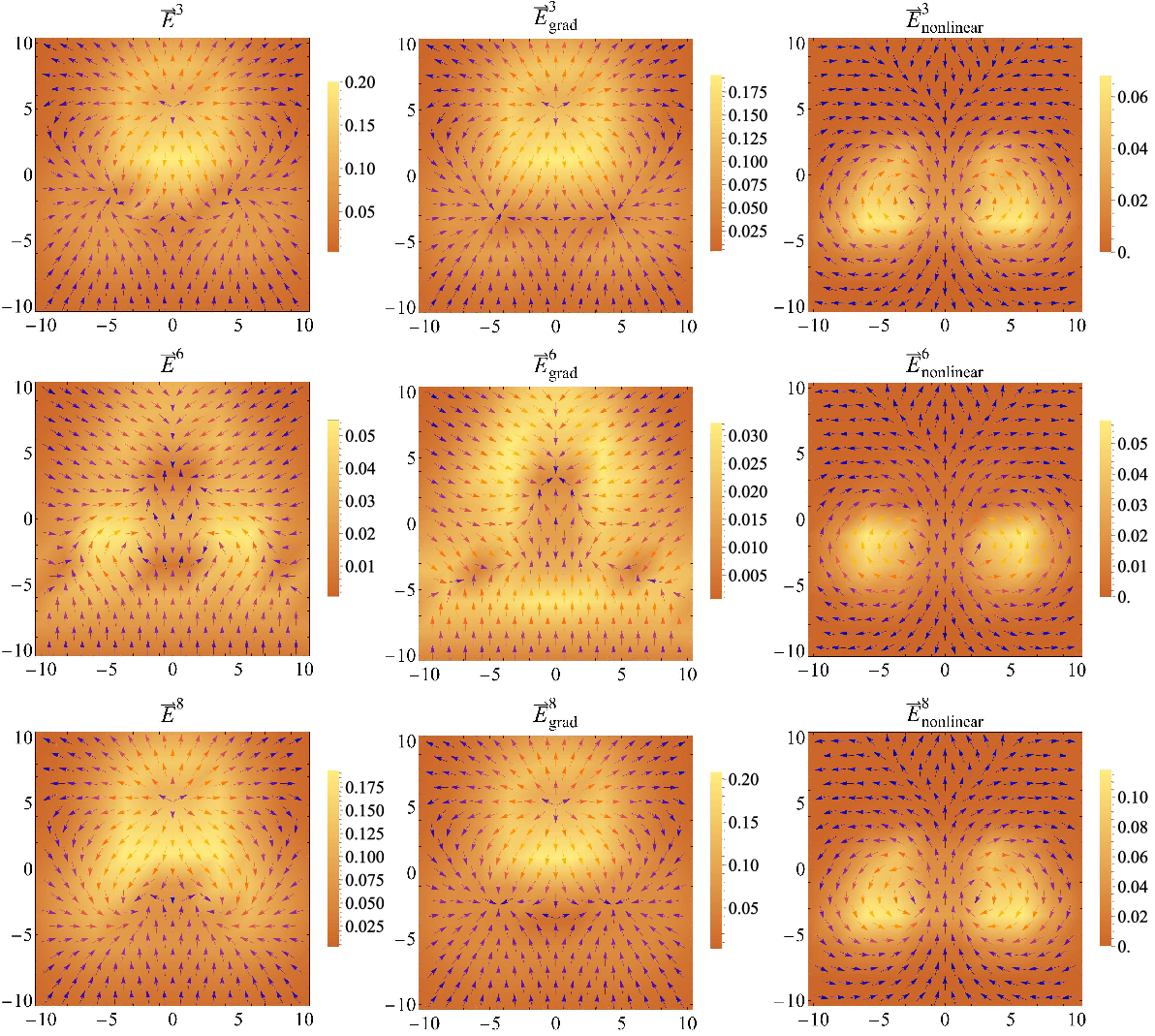}
\caption{Force lines of the electric fields $\vec{E}^{3,6,8}$ in the $z = 0$ plane.}
\label{fig_E_3_6_8}
\end{figure}

\begin{figure}[H]
\centering
\includegraphics[width=1.\linewidth]{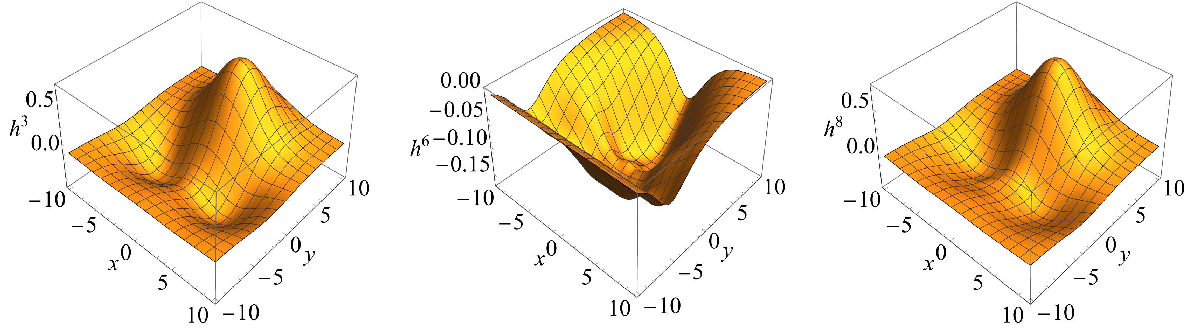}
\caption{Potentials $A^{3,6,8}_t = h^{3,6,8}$ in the $z = 0$ plane.}
\label{fig_pot_3_6_8}
\end{figure}

\begin{figure}[H]
\centering
\includegraphics[width=1.\linewidth]{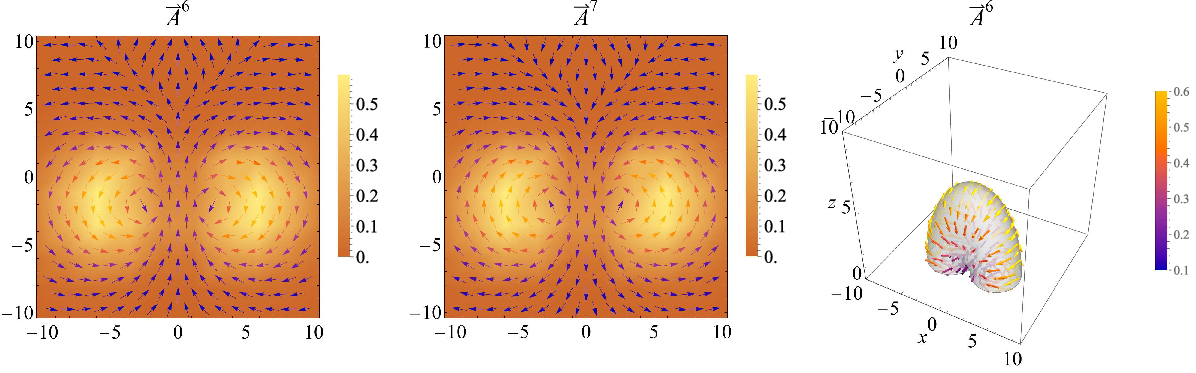}
\caption{Force lines of the vector potentials $\vec{A}^{6,7}$ in the $z = 0$ plane and the spatial distribution of the potential $\vec{A}^{6}$.
}
\label{fig_pot_A_6_7}
\end{figure}

\begin{figure}[H]
\centering
\includegraphics[width=1.\linewidth]{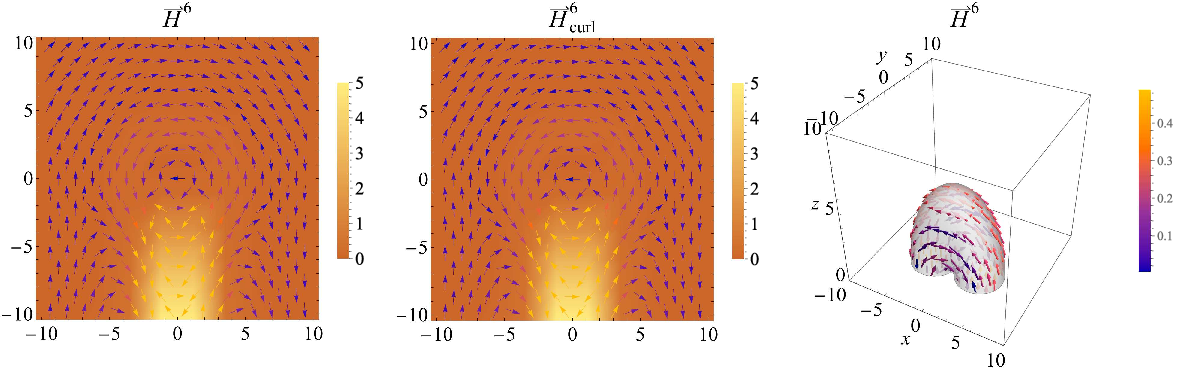}
\caption{Force lines and the curl component of the color magnetic field $\vec{H}^{6}$ in the $y = 0$ plane,
as well as the spatial distribution of  $\vec{H}^{6}$.
}
\label{fig_H_6}
\end{figure}

The numerical solutions of Eqs.~\eqref{eqn_10}-\eqref{eqn_60} are shown in Figs.~\ref{fig_E_3_6_8}-\ref{fig_H_6} 
for the following particular values of the system parameters appearing in the expressions~\eqref{charge_dens_8}-\eqref{current_6_7}: 
$l_0=5, d_0=4, \left( \rho_0\right)_B = \left( \rho_0\right)_C = -1/2 \left( \rho_0\right)_A=-0.5, \rho_1=-0.2, x_0=3, y_0=-2, \alpha=2, \beta=4, j_0=-2, c=x_0$. The top left panel of Fig.~\ref{fig_E_3_6_8} shows the Y-like distribution of force lines of the total electric field $\vec{E}^3$
which has 
two components: the gradient one given by Eq.~\eqref{E_8_grd} (the photon part in the terminology of Ref.~\cite{Bornyakov:2004uv}) and the nonlinear one given by Eq.~\eqref{E_8_nl} (the monopole part in the terminology of Ref.~\cite{Bornyakov:2004uv}). 

The gradient components \eqref{E_8_grd} for the electric fields $\vec{E}^{3,8}$ are shown in Fig.~\ref{fig_E_3_6_8}. They are sourced by the static quarks located at the vertices of the triangle  $ABC$ with the color charge densities $\rho^{3,8}$ that are sources for the components $A^{3,8}_t$ of the non-Abelain potential $A^{3,8}_\mu$ in Eqs.~\eqref{eqn_10} and \eqref{eqn_50}. 

The nonlinear components  \eqref{E_8_nl} for the electric fields $\vec{E}^{3,8}$ are shown in Fig.~\ref{fig_E_3_6_8}. According to Eq.~\eqref{E_8_nl}, they are sourced by the components $A^{6,7}_t$ and $\vec{A}^{6,7}$ of the vector potential $A^{6,7}_\mu$.

The time components of the potentials $A^{3,6,8}_t = h^{3,6,8}$ are shown in Fig.~\ref{fig_pot_3_6_8}. According to Eq.~\eqref{el_fields}, the gradients of these components of the four-potential form the gradient components of the color electric fields $\vec{E}^{3,6,8}$. 

The components $\vec{A}^{6,7}$ that are needed to create the nonlinear components of the color electric fields are shown in Fig.~\ref{fig_pot_A_6_7}, which also shows the spatial distribution of the potential  $\vec{A}^6$. The results of numerical calculations indicate that $\vec{A}^{3,8} \approx 0$. 

The color magnetic field $\vec{H}^{6}$ and its curl component are shown in Fig.~\ref{fig_H_6}, which also shows the spatial distribution of the magnetic field $\vec{H}^6$. The numerical calculations indicate that $\vec{H}^{3,8} \approx 0$. According to the numerical calculations, the spatial distribution of the magnetic field $\vec{H}^7$ is similar to that of $\vec{H}^6$.

\begin{figure}[t]
    \begin{center}
        \includegraphics[width=.49\linewidth]{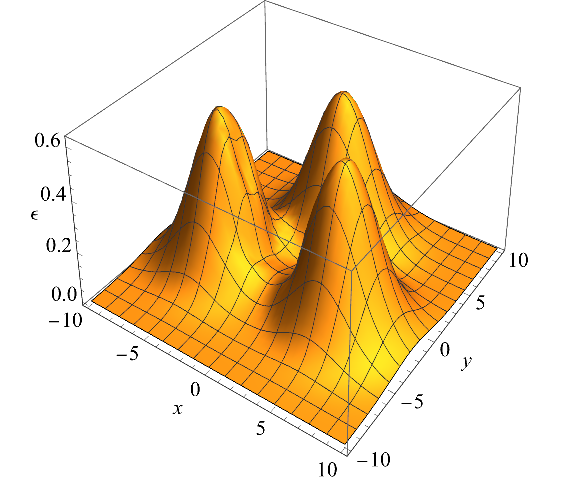}
        \includegraphics[width=.49\linewidth]{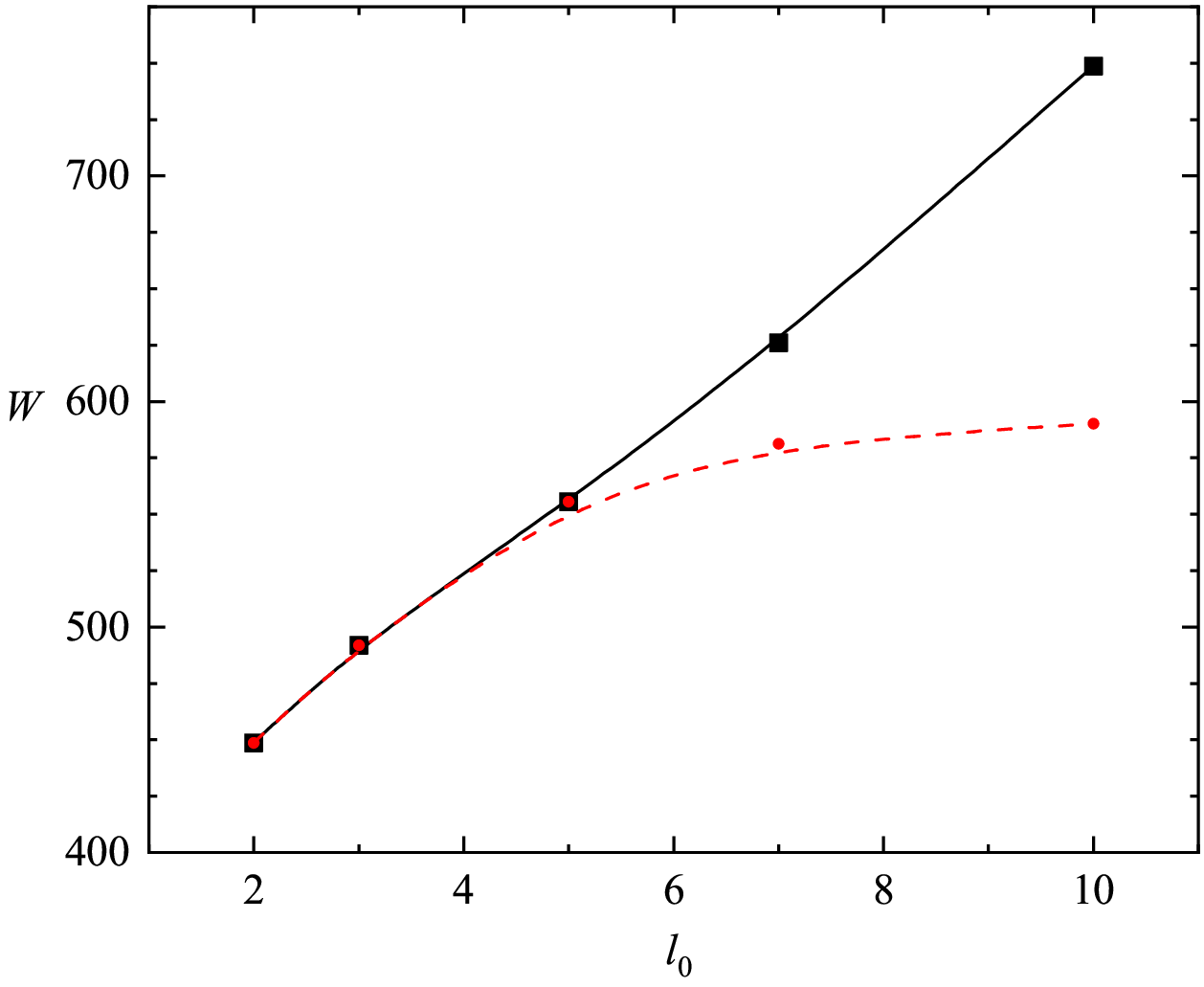}
    \end{center}
    \vspace{-.5cm}
    \caption{ Left panel: the energy density $\varepsilon$ from Eq.~\eqref{energy_dens} for the values of the system parameters given at the beginning of Sec.~\ref{num_res}. Right panel: the total energy $W$ from Eq.~\eqref{full_en} as a function of~$l_0$. The red dashed line is plotted for fixed geometric parameters $x_0, y_0, \alpha, \beta$, and the black solid line~-- for variable parameters.}
     \label{full_energy}
\end{figure}

One of the purposes of the present paper is to compare the total energy of the field configurations obtained with the potential of the three-quark configuration found in lattice calculations. In Ref.~\cite{Sakumichi:2015rfa}, the Abelian part of the 3Q potential has the functional form
\begin{equation}
	V_{3Q}^{\text{Abel}} = \sigma^{\text{Abel}}_{3Q} L_{\text{min}} 
	- \frac{A^{\text{Abel}}_{3Q}}{R} + C_{3Q}^{\text{Abel}} ,
\label{static_pot_1}
\end{equation}
where $1/R = \sum_{i < j} 1/\left| \vec{r}_i - \vec{r}_j\right|$, $\vec{r}_i$ are coordinates of the quarks. 
This form was first presented in lattice QCD in Refs.~\cite{Takahashi:2000te,Takahashi:2002bw}. 

In our study, the quarks are located at the vertices  of an equilateral triangle; for this reason, we compare  the results obtained here with the results of Ref.~\cite{Bornyakov:2004uv} where the quarks are located at the vertices  of an equilateral triangle, and therefore the expression~\eqref{static_pot_1} is modified to
\begin{equation}
	V^{3Q}\left( L_Y\right) = V^{3Q}_0 - 3 \sqrt{3} \frac{\alpha^{3Q}}{L_Y} 
	+ \sigma^{3Q} L_Y ,
\label{static_pot}
\end{equation}
where $V^{3Q}_0, \alpha^{3Q}$, and $\sigma^{3Q}$ are some parameters. $L_Y$ is a characteristic size in the given physical system;
in our case it coincides in order of magnitude with $l_0$ (the length of the side of the triangle $ABC$). 
The second term on the right-hand side of the expression~\eqref{static_pot} describes the behavior of the static potential at small distances between quarks, and the third term corresponds to the confining potential. Note that the expression for $V^{3Q}$ contains monopole and photon parts, and the monopole term increases linearly for quite large $L_Y$, whereas the photon component remains approximately constant.

For comparison with those results, using the expression for the energy density from Eq.~\eqref{energy_dens},
we calculate the profile of the total energy of the  configuration under consideration
\begin{equation}
	W\left( l_0\right)  = 2 \int_{0}^{\infty} \int_{- \infty}^{\infty} \int_{- \infty}^{\infty} \varepsilon(x,y,z) dx dy dz  
\label{full_en}
\end{equation}
as a function of the characteristic size of our configuration $l_0$.
The results of calculations are given in Fig.~\ref{full_energy}. The red dashed line is constructed for fixed geometric parameters describing our system: $l_0$ is a characteristic size of the system consisting of three quarks; $x_0$ and $y_0$ are the positions of the current densities $\vec{j}^{\,6,7}$; $\alpha$ and $\beta$ are parameters determining the profile of the current densities $\vec{j}\,^{6,7}$ (see Appendix~\ref{dens_charges_currents}). It is seen that for rather large values of $l_0$ this profile differs considerably from a linear dependence predicted by lattice calculations. That is because when the characteristic distance between the static quarks increases, all the aforementioned parameters must also change, since actually (or in lattice calculations)  a quantum system consisting of three quarks and of the fields created by them is self-consistent. Such self-consistency means that when the distance between the quarks changes, all other system parameters change as well. In our approach, we do not know how such changes happen, and for this reason we adjust these parameters: the sizes of the triangle, at the vertices of which static sources (quarks) are located, the positions of the current densities, and the parameters determining the profile of the current densities. Notice that in order to have satisfactory agreement with lattice calculations, it is necessary to change the aforementioned geometric parameters; unfortunately, working within our approximate model of nonperturbative quantization, we do not know the law according to which these parameters are being changed. For this reason, we have to choose these geometric parameters arbitrarily. 

Consistent with this observation, for a quite large  $l_0$, we choose the values of these parameters  in such a way that we have a linear dependence (the black solid curve) of the total energy on the characteristic size $l_0$ between the quarks. In Table~\ref{table_1}, we show the values of these parameters for different $l_0$.

\begin{table}[h!]
\begin{tabular}{|c|c|c|c|c|}
\hline
 $l_0$	&  $x_0$	&  $y_0$	&  $\alpha$& $\beta$ \\
\hline
 2	&  3	&  -2	&  2	&  4	\\
\hline
 3	&  3	&  -2	&  2	&  4	\\
\hline
 5	&  3	&  -2	&  2	&  4	\\
\hline
 7	&  3.1	&  -2	&  2	&  4	\\
\hline
 10	& 7	&  -4	&  8	&  4	\\
\hline
\end{tabular}
\caption{The parameters $x_0, y_0, \alpha$, and $\beta$ for different $l_0$ of the black solid curve in the right panel of Fig.~\ref{full_energy}.}
\label{table_1}
\end{table}

Summarizing the results obtained,
\begin{enumerate}
\item According to the results shown in Fig.~\ref{fig_E_3_6_8}, the color electric fields $\vec{E}^{3,8}$ have two well-distinguishable components~-- the gradient and nonlinear ones. The first component $\vec{E}^{3,8}_{\text{grad}}$  is sourced by the color charges $\rho^{3,8}$ on the right-hand sides of Eqs.~\eqref{eqn_10} and \eqref{eqn_50}. These charges are located at the vertices of the triangle  $ABC$, and their densities are given by the expressions~\eqref{charge_dens_8}. According to Eq.~\eqref{E_8_nl}, the nonlinear term $\vec{E}^{3,8}_{\text{nonlinear}}$ is created by the scalar potentials $h^{6,7}$ and the vector potentials $\vec{A}^{6,7}$ (see also the item~\ref{A_6_7} below). 
\item Comparison of the profiles of the color electric fields shown in Fig.~\ref{fig_E_3_6_8} reveals satisfactory agreement of our results with the results obtained in lattice calculations. In both cases, there are: 
\begin{enumerate}
\item the Y-like distribution for the total color electric fields $\vec{E}^{3,8}$, see the left panels in the top and bottom rows of Fig.~\ref{fig_E_3_6_8}; 
\item the Y-like distribution for the gradient component of the electric field (the photon-like component in terms of lattice calculations of Ref.~\cite{Bornyakov:2004uv}), compare the middle panels in the top and bottom rows of our Fig.~\ref{fig_E_3_6_8} and  Figure~9 of Ref.~\cite{Bornyakov:2004uv}; 
\item the curl component of the electric field (the monopole-like component in terms of lattice calculations of Ref.~\cite{Bornyakov:2004uv}), compare the right panels in the top and bottom rows of our Fig.~\ref{fig_E_3_6_8} and  Figure~9 of Ref.~\cite{Bornyakov:2004uv}. 
\end{enumerate}
\item Since the densities of the color charges $\rho^{3,8}$ have the same profile, the electric fields $\vec{E}^{3,8}$ have the same structure.
\item There are no sources for the potentials $h^{6,7}$ in Eq.~\eqref{eqn_30}, and therefore the electric field 
in the middle panel in the second row of Fig.~\ref{fig_E_3_6_8} has no sources. This field is created by the nonlinear terms in Eq.~\eqref{eqn_30}.
\item 
\label{A_6_7}
The lattice calculations of Ref.~\cite{Bornyakov:2004uv} indicate that the color electric field has a monopole component sourced by a solenoidal current. In order to have such a component in our calculations, it is necessary to have solenoidal current densities $\vec{j}^{6,7}$ in Eq.~\eqref{eqn_40}; for this purpose, we use the expression~\eqref{current_6_7} given in Appendix~\ref{dens_charges_currents}. 
\item The magnetic fields $\vec{H}^{6,7}$ have a toroidal structure when the force lines are basically concentrated inside the solenoid created by the currents $\vec{j}^{6,7}$, 
see the right panel of Fig.~\ref{fig_H_6}. 
\item The magnetic fields $\vec{H}^{3,8} \approx 0$. This result has the explanation that, according to Eq.~\eqref{magn_fields}, these fields have two components: the curl one, 
$\vec{H}^{3,8}_{\text{curl}} = \curl{\vec{A}^{3,8}}$, and the nonlinear one,
$\vec{H}^{3,8}_{\text{nonlinear}} \propto \left[ \vec{A}^6 \times \vec{A}^7 \right]$; since
$\vec{A}^{3,8} \approx 0$, and the vectors  $\vec{A}^6$ and $\vec{A}^7$ are parallel, both these components are zero vectors. 
\item The profile of the energy density shown in Fig.~\ref{full_energy} (left panel) resembles (at least qualitatively) the profile of the Abelian action density shown in Figures 10 and 12 of Ref.~\cite{Bornyakov:2004uv}. 
\item We have obtained the profile of the total energy as a function of the characteristic size of the configuration under consideration. It is shown that this profile coincides qualitatively with the behavior of the photon component of the static potential obtained in lattice calculations of Ref.~\cite{Bornyakov:2004uv}. 
\end{enumerate}

\section{Discussion of the connection with lattice calculations}
\label{dis_con}

The results of calculations presented in Sec.~\ref{Y_distribution} indicate that there is at least a qualitative agreement with the results of lattice calculations: 
(a)~we have obtained the Y-like distribution of the color electric field with the sources in the form of static quarks located at the vertices of the triangle; 
(b)~there is the Y-like component of the gradient color electric field; (c)~there is the curl component of this field. 

It enables us to assume that there may exist some connection between nonperturbative QCD, for which the results can be obtained in lattice calculations, and Yang-Mills-Proca theory under consideration. We assume that  Yang-Mills-Proca theory may serve as some approximation to nonperturbative QCD as follows.
Let us consider a SU(3) gluon condensate varying in space,
\begin{equation}
	F^2\left( x^\alpha \right) = \left\langle 
		\hat{F}^a_{\mu \nu} \left( x^\alpha \right) \hat{F}^{a \mu \nu} \left( x^\alpha \right) 
	\right\rangle = 
	\left\langle 
			\hat{F}^{\bar{a}}_{\mu \nu} \left( x^\alpha \right) \hat{F}^{\bar{a} \mu \nu} \left( x^\alpha \right) 
		\right\rangle 
	+ \left\langle 
				\hat{F}^m_{\mu \nu} \left( x^\alpha \right) \hat{F}^{m \mu \nu} \left( x^\alpha \right) 
			\right\rangle 
\label{gluon_cond}
\end{equation}
with
\begin{equation}
\begin{split}
	\hat{F}^{\bar{a}}_{\mu \nu} & = \hat{\mathcal{F}}^{\bar{a}}_{\mu \nu} 
	+ g f^{\bar{a} m n} \hat{A}^m_\mu \hat{A}^n_\nu , 
\\ 
	\hat{F}^m_{\mu \nu} & = \hat{\mathfrak{F}}^m_{\mu \nu} + g f^{\bar{a} m n} \left( 
		\hat{A}^n_\mu \hat{A}^{\bar{a}}_\nu - \hat{A}^{\bar{a}}_\mu \hat{A}^n_{\nu} 
	\right) .
\end{split}
\label{filed_strenth_tensors}
\end{equation}
Here $a = 1,2, \ldots, 8, \bar{a} = 3,6,7,8$, and $m = 1,2,4,5$; 
 $\hat{A}^{\bar{a}}_\mu$ belongs to a subalgebra spanned on the generators $\lambda^{3, 6, 7, 8}$; 
$
	\hat{\mathcal{F}}^{\bar{a}}_{\mu \nu} = \partial_\mu \hat{A}^{\bar{a}}_\nu 
	- \partial_\nu \hat{A}^{\bar{a}}_\mu +
	g f^{{\bar{a}} {\bar{b}} {\bar{c}}} \hat{A}^{\bar{b}}_\mu \hat{A}^{\bar{c}}_\nu
$; 
$
	\hat{\mathfrak{F}}^m_{\mu \nu} = \partial_\mu \hat{A}^m_{\nu} - \partial_\nu \hat{A}^m_{\mu}
$. 
We assume that in some physical situations all degrees of freedom  $A^a_\mu$ are divided into two different in principle sets. This happens, for example, when a flux tube forms between two quarks, and there exists a longitudinal electric field between  these quarks. A similar situation occurs for a system of three quarks. In both situations there are nonzero quantum averages of color electric fields that are described using Eq.~\eqref{0_20}. The first set contains ``almost classical'' degrees of freedom in the sense that $\left\langle \hat{A}^{3,6,7,8}_\mu \right\rangle \approx A^{3,6,7,8}_\mu$. The second set contains ``purely quantum'' degrees of freedom in the sense that
$\left\langle \hat{A}^{1, 2, 4, 5}_\mu \right\rangle \approx 0$. Physically, this means that, in some approximation in some processes in QCD, there are almost classical, $A^{3,6,7,8}_\mu$, and purely quantum, $A^{1,2,4,5}_\mu$, degrees of freedom. 

Using Eq.~\eqref{filed_strenth_tensors}, the expression for the gluon condensate~\eqref{gluon_cond} takes the form
\begin{equation}
\begin{split}
	F^2 = & \mathcal{F}^{\bar{a}}_{\mu \nu} \mathcal{F}^{\bar{a} \mu \nu} 
	+ 2 g f^{\bar{a} m n} \mathcal{F}^{\bar{a}}_{\mu \nu} 
	\left\langle \hat{A}^{m \mu} \hat{A}^{n \nu} \right\rangle 
	+ g^2 f^{\bar{a} m n} f^{\bar{a} p q} 
	\left\langle 
		\hat{A}^{m}_\mu \hat{A}^{n}_\nu \hat{A}^{p \mu} \hat{A}^{q \nu}
	\right\rangle 
\\ 
	& + \left\langle  \mathfrak{F}^m_{\mu \nu} \mathfrak{F}^{m \mu \nu} 
	\right\rangle 
	+ 2 g f^{\bar{a} m n} 
	\left( 
		\left\langle 
			\hat{\mathfrak{F}}^m_{\mu \nu} \hat{A}^{n \mu} 
		\right\rangle A^{\bar{a} \nu} 
		- \left\langle 
			\hat{\mathfrak{F}}^m_{\mu \nu} \hat{A}^{n \nu} 
		\right\rangle A^{\bar{a} \mu}
	\right) 
\\
	& + 2 g^2 f^{m n \bar{a}} f^{m p \bar{b}} 
	\left( 
		\left\langle \hat{A}^n_\mu \hat{A}^{p \mu} \right\rangle A^{\bar{a}}_\nu A^{\bar{b} \nu} 
		- \left\langle \hat{A}^n_\mu \hat{A}^{p \nu}\right\rangle A^{\bar{a}}_\nu A^{\bar{b} \mu}
	\right) . 
\end{split}
\label{gluon_condensate_2}
\end{equation}
The gluon condensate \eqref{gluon_condensate_2} is in practice the expectation value of SU(3) Lagrangian. 
Therefore we can employ it to derive field equations for the ``almost classical'' degrees of freedom $A^{\bar{a}}_\mu$
by varying it with respect to $A^{\bar{a}}_\mu$; as a result, we have the equation
\begin{equation}
\begin{split}
	D_\nu \mathcal{F}^{\bar{a} \mu \nu} 
	+ g f^{\bar{a} m n} \left( 
	\partial_\nu \left\langle \hat{A}^{m \mu} \hat{A}^{n \nu} \right\rangle 
	- \left\langle 
	\hat{\mathfrak{F}}^{m \mu \nu} \hat{A}^n_\nu \right\rangle 
	\right) 
	+ \frac{g^2}{2} f^{m n \bar{a}} f^{m p \bar{b}} 
	\left( 
		\left\langle \hat{A}^n_\nu \hat{A}^{p \mu}\right\rangle A^{\bar{b \nu}} 
		- \left\langle \hat{A}^n_\nu \hat{A}^{p \nu}\right\rangle A^{\bar{b} \mu}
	\right) = 0. 
\end{split}
\label{field_eqn}
\end{equation}
Two last terms in \eqref{field_eqn} can be rewritten in the form
\begin{align}
	G^{\bar{a} \bar{b}, \mu \nu} A^{\bar{b}}_\nu & = - \frac{g^2}{2} f^{m n \bar{a}} f^{m p \bar{b}} 
		\left( 
			\left\langle \hat{A}^n_\nu \hat{A}^{p \mu}\right\rangle A^{\bar{b} \nu} 
			- \left\langle \hat{A}^n_\nu \hat{A}^{p \nu}\right\rangle A^{\bar{b} \mu}
		\right) ,
\label{current_1}
\\
	j^{\bar{a} \mu} & = g f^{\bar{a} m n} \left( 
			\partial_\nu \left\langle \hat{A}^{m \mu} \hat{A}^{n \nu} \right\rangle 
			- \left\langle 
			\hat{\mathfrak{F}}^{m \mu \nu} \hat{A}^n_\nu \right\rangle 
	\right) , 
\label{current_2}
\end{align}
and this enables us to rewrite Eq.~\eqref{field_eqn} as
\begin{equation}
\begin{split}
	D_\nu \mathcal{F}^{\bar{a} \mu \nu} 
	- G^{\bar{a} \bar{b}, \mu \nu} A^{\bar{b}}_\nu = - j^{\bar{a} \mu}. 
\end{split}
\label{field_eqn_cond}
\end{equation}
It is seen that Eq.~\eqref{0_20} used by us for calculations can be obtained from Eq.~\eqref{field_eqn_cond}
using the approximation for the mass term,
\begin{equation}
	G^{\bar{a} \bar{b}, \mu \nu} A^{\bar{b}}_\nu  \approx m^2 A^{\bar{a} \mu} , 
\label{approx_1}
\end{equation}
and the expressions  \eqref{charge_dens_8}, \eqref{charge_dens}, and \eqref{current_6_7} for the currents 
$ j^{\bar{a} \mu}$ created by 2- and 4-point Green functions appearing on the right-hand side of the expression~\eqref{current_2}.

Note that using the expression \eqref{gluon_condensate_2}, one can in principle obtain also an equation for a 2-point Green function 
$
	G^{m n}_{\phantom{m n} \mu \nu}\left( x, y\right)  = \left\langle 
		\hat{A}^m_\mu \left( x\right)  \hat{A}^n_\nu\left( y\right) 
	\right\rangle 
$. But in deriving this equation there will arise the following big problems: (a)~this Green function is a function of two variables $x$ and $y$, and this leads to the appearance of two differential equations;
(b)~in these equations, there will be a 4-point Green function
$
	G^{m n p q}_{\phantom{m n p q} \mu \nu \rho \sigma}\left( x, y, z,u\right)  = \left\langle 
		\hat{A}^m_\mu \left( x\right)  \hat{A}^n_\nu\left( y\right) \hat{A}^p_\rho \left( z\right) 
		\hat{A}^q_\sigma \left(u\right) 
	\right\rangle 
$, for which it will be necessary either to write down its own equation or to use some approximation to close these equations; 
(c)~there is an ambiguity  in defining the quantity  $\partial_\nu \left\langle \hat{A}^{m \mu} \hat{A}^{n \nu} \right\rangle$: 
whether it is a derivative of the first operator $\left\langle \partial_\nu \hat{A}^{m \mu} \hat{A}^{n \nu} \right\rangle$ or
it is  a derivative of the second operator $\left\langle \hat{A}^{m \mu} \partial_\nu \hat{A}^{n \nu} \right\rangle$. 
Keeping these problems in mind, in the present study we restrict ourselves to Eq.~\eqref{field_eqn_cond} with the corresponding approximations described above. 

Summarizing, we can say that, \textit{in lattice calculations, a direct numerical quantization of a system is carried out using the Feynman path integral, whereas, in our calculations, we, using an approximation of ``almost classical'' and ``purely quantum'' degrees of freedom, seek an approximate solution for the ``almost classical'' degrees of freedom, while for the ``purely quantum'' degrees of freedom we employ some approximation.
}

\section{Conclusions}

Thus we have shown that, within SU(3) non-Abelian Yang-Mills-Proca theory with external sources in the form of three static quarks, there are solutions  that are in satisfactory agreement with the results of lattice calculations in QCD: there is the Y-like distribution of the color electric field, as well as the curl component of this electric field. The Y-like distribution of the electric field is created by static quarks located  at the vertices of an equilateral triangle.
 
Let us enumerate the main results of the study: 
\begin{itemize}
\item We have obtained regular solutions within Yang-Mills-Proca theory with external sources in the form of:  
(a)~color charges (simulating three static quarks) located at the vertices of the triangle;
 (b)~charges distributed along the sides of this  triangle;
 (c)~solenoidal  currents. 
 \item It is shown that the color electric and magnetic fields have two components: the gradient/curl term for the electric/magnetic field, respectively,  and the nonlinear component is present in both cases. 
 \item It is shown that the distribution of the color electric field adequately represents the results obtained in lattice calculations: this field has a  Y-like distribution, and the curl electric field is also present.
 \item The presence of the Y-like field is caused by the quarks located at the vertices of the triangle. The curl electric field is sourced by the solenoidal  currents.
 \item The color magnetic field has only $\vec{H}^{6,7}$ nonzero components. The components  $\vec{H}^{3,8} \approx 0$, since $\vec{A}^{3,8} \approx 0$ and the potentials $\vec{A}^{6,7}$ are parallel to each other.
 \item We have obtained the energy profile as a function of the characteristic distance between quarks for a system with variable geometric parameters;  this profile  coincides qualitatively with the profile of the photon part of the static potential  obtained in Ref.~\cite{Bornyakov:2004uv}. 
 \item We have shown that using a condensate varying in space, one can obtain the Yang-Mills-Proca equation sourced by quantum averages 
 $
 		\left\langle 
 			\hat{A}^m_\mu \left( x\right)  \hat{A}^n_\nu\left( y\right) 
 		\right\rangle $, as well as there are mass terms of the form $G^{a b, \mu \nu} A^b_\nu$. 
\end{itemize}

We began studies in this direction in Ref.~\cite{Dzhunushaliev:2024dzp} where we examined a flux tube within
non-Abelian Proca theory with external sources simulating two quarks creating such a tube. The results obtained in Ref.~\cite{Dzhunushaliev:2024dzp} and in the present paper indicate that in both cases the field energy of the soliton-like objects found increases  with increasing distance between the sources creating
the corresponding non-Abelian field. This means that in both cases there is the same mechanism ensuring the interaction potential between the sources which grows with their separation. 

Most unsolved problems in QCD are related to the problem of nonperturbative quantization. In Ref.~\cite{heis}, W.~Heisenberg suggested the idea of nonperturbative quantization based on an infinite set of differential Dyson-Schwinger equations for all Green functions; for solving this set, Feynman diagrams are not used. Later (see, e.g., Refs.~\cite{Bender:1999ek,Frasca,Dzhunushaliev:2022apb}), this idea was used for nonperturbative quantization in QCD. For practical application of this approach, the infinite set of equations is to be cut off to a finite set. In the present study we avoid this problem by using a gluon condensate, which is a function of spatial coordinates and simultaneously is a Lagrangian density. Varying the corresponding action, we derived the Yang-Mills-Proca equation whose solution can be found using some approximations for a 2-point Green function. For further advance in this direction it is necessary to obtain an equation for the aforementioned Green function, but in deriving it one faces the problems described in Sec.~\ref{dis_con}, as well as it is necessary to take into consideration spinor fields (which describe quarks).

Notice that a comparison of the results obtained in lattice calculations and the results obtained in the direction presented here may lead to a considerable improvement in understanding of the physical processes  occurring  in nonperturbative QCD.

We note in conclusion that the present study is one of the first steps  towards the use of nonperturbative quantization as suggested by W.~Heisenberg in Ref.~\cite{heis} for quantising a nonlinear spinor field.  One may expect that this direction will be effective in estimating hadron and glueball masses, in understanding spontaneous chiral symmetry breaking and axial U(1) anomaly. But of course this can be done only after inclusion of an equation for a 2-point Green function $G^{m n}_{\phantom{m n} \mu \nu}$, 
as well as of equations describing 
$
	\left\langle \hat{\bar{\psi}} \hat{A}^a_\mu \hat{\psi}\right\rangle 
$. It is possible that to make progress in this direction it will be necessary to use first $n$ equations of a set of the Dyson-Schwinger equations with $n > 2$. But this is a task for further investigations in this direction.

\section*{Acknowledgements}

We gratefully acknowledge support provided by the program No.~AP26195069 (Bound states in Maxwell, Yang-Mills, Proca theories with and without gravity in the presence of spinor fields) of the Committee of Science of the Ministry of Science and Higher Education of the Republic of Kazakhstan. 
We thank  V.~Braguta for valuable consultations. 

\appendix

\section{Equations for numerical calculations}
\label{full_eqns}

For the potentials \eqref{gauge_pot}, the field equations \eqref{0_20} take the form
\begin{align}
	&
	\laplacian{h^3} - m^2 h^3 
	- \frac{g}{2} \div{\left( h^6 \vec{A}^7 - h^7 \vec{A}^6 \right)} 
	- \frac{g}{2}\left( 
		- \vec{A}^6 \cdot \vec{E}^7 + \vec{A}^7 \cdot \vec{E}^6 
	\right) 
	= \rho^{3} , 
\label{eqn_10} \\
	&
	\left( \curl{\curl{\vec{A}^{\,3}}}\right)_i + m^2 A^3_i 
	- \frac{g}{2} \partial_j \left( A^6_i A^7_j - A^7_i A^6_j\right) 
	- \frac{g}{2} \left( 
		h^6 E^7_i - \left[\vec{A}^{\,6} \times \vec{H}^7\right]_i  
		- h^7 E^6_i + \left[\vec{A}^{\,7} \times \vec{H}^6 \right]_i
	\right) 
	= j^{3}_i , 
\label{eqn_20} \\
	& 
	\laplacian{h^{6, 7}} - m^2 h^{6, 7} 
	+ \frac{g}{2} \div{ \left[ 
		\left( h^3 - \sqrt{3} h^8\right) \vec{A}^{\,7, 6} - 
		h^{7, 6} \left( \vec{A}^{\,3} - \sqrt{3} \vec{A}^{\,8}\right) 
	\right] }
\nonumber \\
	&
	+	\frac{g}{2} \left[ 
		\vec{A}^{\,7, 6} \cdot \left( \vec{E}^3 - \sqrt{3} \vec{E}^8 \right) 
		+ \vec{E}^{7, 6} \cdot \left( - \vec{A}^{\,3} + \sqrt{3} \vec{A}^{\,8} \right) 
	\right]  
	= \rho^{6, 7} , 
\label{eqn_30}\\
	&
	\left(\curl{\curl{\vec{A}^{6, 7}}}\right)_i + m^2 A^{6, 7}_i 
	+ \frac{g}{2} \partial_j
	\left[ 
		\left( A^3_i - \sqrt{3} A^8_i\right) A^{7, 6}_j 
		- A^{7, 6}_i \left(A^3_j - \sqrt{3} A^8_j\right) 
	\right] 
\nonumber \\
	& 
	+ \frac{g}{2} \left\{ 
			\left( h^3 - \sqrt{3} h^8 \right) E^{7, 6}_i 
			- h^{7, 6} \left( E^8_i - \sqrt{3} E^3_i \right) 
		+ \left[\vec{A}^{\,7, 6} \times \left( \vec{H}^3 - \sqrt{3} \vec{H}^8 \right) \right]_i
		- \left[\left( \vec{A}^{\,3} - \sqrt{3} \vec{A}^{\,8} \right) \times \vec{H}^{7, 6}\right]_i
	\right\} = j^{6, 7}_i , 
\label{eqn_40}\\
	&
	\laplacian{h^8} - m^2 h^8 + 
	\frac{g}{2} \div{ 
	\left[ 
		\vec{A}^{\,6} \left( h^3 - \sqrt{3} h^8\right) 
		- h^6 \left( \vec{A}^{\,3} - \sqrt{3} \vec{A}^{\,8}\right) 
	\right] 
	} 
	- \frac{\sqrt{3}}{2} g \left( 
		\vec{A}^{\,6}\cdot \vec{E}^7 - \vec{A}^{\,7}\cdot \vec{E}^6
	\right) 
	= \rho^{8} , 
\label{eqn_50}\\
	&
	\left( \curl{\curl{\vec{A}^{\,8}}}\right)_i + m^2 A^{8}_i 
	+ \frac{\sqrt{3}}{2} g \,\partial_j \left( A^6_i A^7_j - A^7_i A^6_j \right)  
	+\frac{\sqrt{3}}{2} g \left( 
		h^6 E^7_i - h^7 E^6_i 
		- \left[ \vec{A}^{\,6} \times \vec{H}^7\right]_i 
		+ \left[ \vec{A}^{\,7} \times \vec{H}^6\right]_i 
	\right) 
	= j^{8}_i ,
\label{eqn_60}
\end{align}
where $\vec{A}^a = \left\lbrace v^a, u^a, w^a \right\rbrace $.

The components of Eq.~\eqref{divergence} have the following explicit form 
(henceforth we consider the case where  $j_z^{3,8} = 0$ but  $j_z^{6,7} \neq 0$):
\begin{align}
	m^2  \div{\vec{A}^{\,3}} = & 
	\div{\vec{j}^{\,3}} - \frac{g}{2} 
	\left( 
		A^7_t \rho^{6 }- \vec{A}^{\,7} \cdot\vec{j}^{\,6 } - A^6_t \rho^{7} + \vec{A}^{\,6} \cdot\vec{j}^{\,7} 
	\right) , 
\label{constraint_10}\\
	m^2  \div{\vec{A}^{\,6, 7}}= &  \div{\vec{j}^{\,6, 7}} 
	\mp \frac{g}{2}\left[ 
		A^3_t \rho^{7, 6 }- \vec{A}^{\,3} \cdot \vec{j}^{\,7, 6} - A^{7, 6}_t \rho^{3 } + \vec{A}^{\,7, 6} \cdot \vec{j}^{\,3} 
			\right. 
\nonumber \\
	&
	\left. 
				+ \sqrt{3} \left( A^{7, 6}_t \rho^{8 }-\vec{A}^{\,7, 6} \cdot \vec{j}^{\,8 } -A^8_t \rho^{7,6 } + \vec{A}^{\,8} \cdot \vec{j}^{\,7, 6 } \right) 
	\right] , 
\label{constraint_20}\\
	m^2  \div{\vec{A}^{\,8}} = & 
	 \div{\vec{j}^{\,8}} +  \frac{\sqrt{3}}{2} g 
		\left( 
			A^7_t \rho^{6 }- \vec{A}^{\,7} \cdot\vec{j}^{\,6}  - A^6_t \rho^{7 } + \vec{A}^{\,6} \cdot\vec{j}^{\,7} 
		\right). 
\label{constraint_30}
\end{align}
In Eq.~\eqref{constraint_20}, the choice of the sign ``$-$'' in front of the square brackets corresponds to the equation for the divergence of $\vec{A}^{\,6}$,
and the choice of the sign ``+'' --~to the equation for the divergence of $\vec{A}^{\,7}$.
To get rid of mixed derivatives in Eqs.~\eqref{eqn_20}, \eqref{eqn_40}, and \eqref{eqn_60}, it is necessary to use the known relation
$\curl{\curl{\vec{A}^{\,3,6,7,8}}} = \grad{\left( \div{\vec{A}^{\,3,6,7,8}}\right) } - \laplacian{\vec{A}^{\,3,6,7,8}}$, as well as the relations~\eqref{constraint_10}-\eqref{constraint_30}. 
After such substitution, all the equations~\eqref{eqn_10}-\eqref{eqn_60} are of the elliptic type, and they can be numerically solved using the CESDSOL  library. 
It is worth pointing out, however, that after such procedure the resulting equations contain derivatives of the sources, $\partial_\mu j^b_\nu$. 

\section{Densities of color charges and currents
}
\label{dens_charges_currents}

The charge densities $\rho^{3,8}_{A, B, C}$ of static quarks located at the vertices of the triangle $ABC$ (see Fig.~\ref{triangleABC}) are 
\begin{equation}
	\rho^{3,8}_{A, B, C} =  \left( \rho_0\right)_{A, B, C} 
	\exp{
	- \frac{\left(x - x_{A, B, C}\right)^2}{d_0^2} 
	- \frac{\left(y -y_{A, B, C}\right)^2}{d_0^2}  
	- \frac{\left(z - z_{A, B, C}\right)^2}{d_0^2} 
	} .
\label{charge_dens_8}
\end{equation}
Here $ \left( \rho_0\right)_{A, B, C}$ are three parameters describing the magnitude of charges located  at the vertices of the triangle $ABC$: 
$\left( \rho_0\right)_B = \left( \rho_0\right)_C = -1/2 \left( \rho_0\right)_A$; $d_0$ is a characteristic size of the distribution of these charges; $(x, y, z)_{A, B, C}$ 
are $x, y, z$-coordinates of the vertices $A, B, C$. These charge densities serve as sources in Eqs.~\eqref{eqn_10}  and~\eqref{eqn_50}: 
\begin{equation}
	\rho^{3, 8} = \rho^{3, 8}_A + \rho^{3, 8}_B + \rho^{3, 8}_C . 
\label{charge_dens_8_1}
\end{equation} 

\begin{figure}[H]
    \begin{center}
        \includegraphics[width=.4\linewidth]{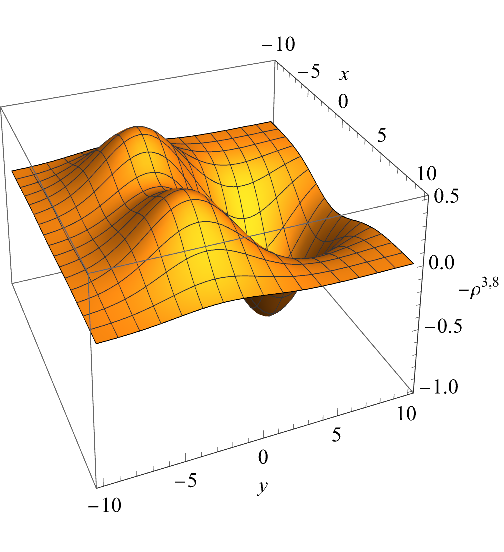}
        \includegraphics[width=.43\linewidth]{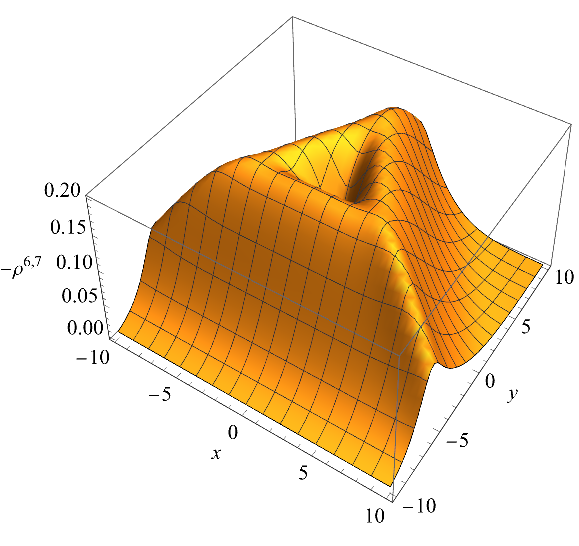}
    \end{center}
    \vspace{-.5cm}
    \caption{The profiles of the color electric charge densities $\rho^{3}=\rho^{8}$ and  $\rho^{6}=\rho^{7}$ in the $z = 0$ plane.
    }
    \label{rho_3_8_6_7}
\end{figure}

Analytical expressions for the densities $\rho^{6,7}$ are
\begin{align}
	\rho^{6,7} = & \rho_{AB} +\rho_{BC} + \rho_{CA} = 
	\frac{\rho_1}{4} \left[ \tanh (x)+1\right] 
	\left[\tanh \left(\frac{x}{\sqrt{3}}+y\right)+1\right] 
	\nonumber \\
	&
	\exp \left\lbrace 
	-\frac{\left(x - \frac{- b_{AB} k_{AB} + k_{AB} y+x}{k_{AB}^2+1}\right)^2 
		+ \left[
		-\frac{k_{AB} \left(-b_{AB} k_{AB}+k_{AB} y+x\right)}{k_{AB}^2+1}-b_{AB}+y
		\right]^2+z^2}{d_0^2}
	\right\rbrace + 
	\nonumber \\
	&
	\frac{\rho_1}{4} \left[\tanh \left(\frac{x}{\sqrt{3}}-y\right)+1\right] 
	\left[1-\tanh \left(\frac{x}{\sqrt{3}}+y\right)\right] 
	\exp{-\frac{\left(y-b_{\text{BC}}\right)^2+z^2}{d_0^2}} + 
	\nonumber \\
	& 
	\frac{\rho _1}{4}  \left[1-\tanh (x)\right] \left[1-\tanh \left(\frac{x}{\sqrt{3}}-y\right)\right] 
	\nonumber \\
	& 
	\exp{
		-\frac{\left(
			x-\frac{-b_{AC} k_{AC} + k_{AC} y+x}{k_{AC}^2+1}
			\right)^2 
			+ \left[
			-\frac{k_{AC} \left(-b_{AC} k_{AC} + k_{AC} y + x \right)}{k_{AC}^2+1}-b_{AC}+y
			\right]^2 + z^2}{d_0^2}
	} . 
\label{charge_dens}
\end{align}
Here $\rho_{AB}$ is the density of a color electric charge located on the side $AB$ of the triangle $ABC$ 
(similarly, for $\rho_{BC}$ and $\rho_{CA}$). The equation $- b_{AB} k_{AB} + k_{AB} y+x = 0$ is an equation of a straight line perpendicular to the side $AB$
which passes through the vertex $A$. An equation of a straight line passing through the points $A$ and $B$ (the side $AB$)
 is  $y = k_{AB} x + b_{AB}$. An equation for a straight line perpendicular to the side $AB$
which passes through the vertex $B$ is written similarly. 
An equation for a straight line for the side  $BC$ is $y = b_{BC}$. The factor with the exponent is chosen so that the charge density
decreases exponentially in the direction perpendicular to the side $AB$.  The factors
$\left[ \tanh (x)+1\right]$ and $\left[\tanh \left(\frac{x}{\sqrt{3}}+y\right)+1\right] $ are chosen so as to approximate a step function
restricting $\rho_{AB}$ by the straight lines passing through the vertices $A$ and $B$ (similarly, for $\rho_{BC}$ and $\rho_{CA}$). 
For better understanding of the distributions of the densities \eqref{charge_dens_8_1} and \eqref{charge_dens}, 
Fig.~\ref{rho_3_8_6_7}  shows the profiles of $\rho^{3,8}$ and $\rho^{6,7}$.

\begin{figure}[H]
\centering
\includegraphics[width=0.5\linewidth]{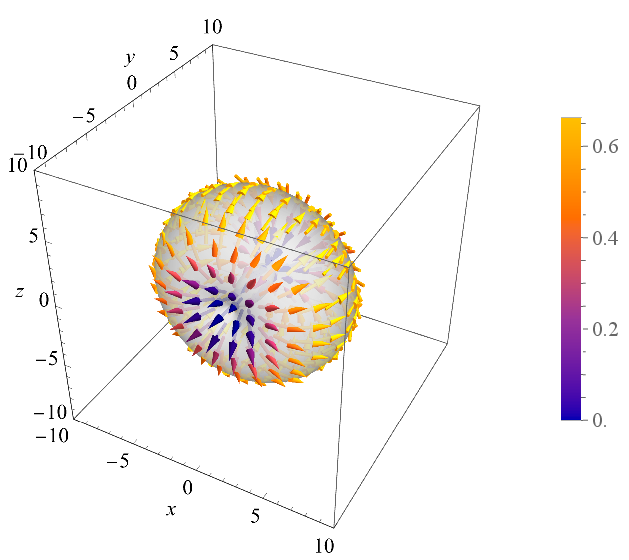}
\caption{The profile of the current $\vec{j}^{\,6}$ given by Eq.~\eqref{current_6_7}.}
\label{solenoid_currnt}
\end{figure}

For the solenoidal current densities $\vec{j}^{\,6, 7}$, we employ some functions that are needed on transforming to toroidal coordinates.
The necessary details about toroidal coordinates $\left(\tau, \sigma, \varphi\right)$ are given in Appendix~\ref{tor_coords}. 
Using the vector~\eqref{tang_vect} tangent to the torus, we introduce the solenoidal current density $\vec{j}^{\,7} = - \vec{j}^{\,6}$ as follows:
\begin{equation}
	\vec{j}^{\,6} = j_0 \vec{h}^\varphi 
	\left[
			- \frac{\beta}{\bar{\tau}} + \frac{\beta}{\bar{\tau}\left(x_0,y_0,0\right)}
		\right] 
	\exp \left[
		- \frac{\beta}{\bar{\tau}} + \frac{\beta}{\bar{\tau}\left(x_0,y_0,0\right)}
	\right] ,
\label{current_6_7}
\end{equation}
where 
$
	\bar{\tau} = \left[ 
		1 - e^{- 2\tau\left( x, y - y_0, z\right) }
	\right]^\alpha
$, $x_0 = c$, and the function $\tau(x, y, z)$ is given by the expression \eqref{tor_tau}. The parameters $x_0$ and $y_0$ specify the location of the circle  $x^2 + z^2 = x_0^2$ in the $y = y_0$ plane; 
$j_0$ is a parameter specifying the magnitude of the current density, and $\alpha$ and $\beta$ are some parameters controlling the profile of the solenoidal current.
The profile of this current on a torus with some $\tau = \text{const}$ is exemplified by Fig.~\ref{solenoid_currnt}.

\section{Toroidal coordinates}
\label{tor_coords}

The transformation from toroidal coordinates $(\tau, \sigma, \varphi)$ to Cartesian coordinates $(x, y, z)$ is performed using the  formulas 
\begin{align}
	\tau = & \log \left(
		\sqrt{\frac{\left(\sqrt{x^2+z^2}+c\right)^2+y^2}{\left(\sqrt{x^2+z^2}-c\right)^2+y^2}}
	\right) , 
\label{tor_tau}\\
	\sigma = & \arccos\left(
		\frac{-c^2+x^2+y^2+z^2}{\sqrt{\left(-c^2+x^2+y^2+z^2\right)^2+4 c^2 y^2}}
	\right) , 
\label{tor_sigma}\\
	\varphi = & \arctan \left(\frac{z}{x}\right) ,
\label{tor_phi}
\end{align}
where the parameter $c$ specifies the location of the circle  $x^2 + z^2 = c^2, y = 0$. The expression $\tau=\text{const}$ specifies the equation of a torus.

The square of the linear element is given by the expression
$$
	dl^2 = \frac{c^2}{\left( \cosh \tau - \cos \sigma \right)^2} 
	\left( 
		d \tau^2 + d \sigma^2 + \sinh^2 \tau d \varphi^2 
	\right) 
	= h^a_{\phantom{a}i} h_{a j} dx^i dx^j , 
$$
where $i, j$ are spatial indices and $a$ is a triad index. The triad  
\begin{equation}
	h^a_{\phantom{a} i} = \frac{1}{\cosh \tau- \cos \sigma} 
	\begin{pmatrix}
		\partial_x \tau & \partial_y \tau & \partial_z \tau \\
		\partial_x \sigma & \partial_y \sigma & \partial_z \sigma \\
		\sinh \tau \partial_x \varphi & \sinh \tau \partial_y \varphi & \sinh \tau \partial_z \varphi
	\end{pmatrix} . 
\label{tangent_vect}
\end{equation}
The unit vector 
\begin{equation}
	h^3_{\phantom{3} i} = h^\varphi_{\phantom{\varphi} i} = N 
	\left\lbrace 
		-4 x \left( y - y_0 \right) , 2 \left(x^2-x_0^2-\left( y - y_0 \right)^2+z^2\right) , -4 z \left( y - y_0 \right)
	\right\rbrace 
\label{tang_vect}
\end{equation}
is tangent to a torus in the $\varphi = \text{const}$ plane, and 
$$
	N = \left\lbrace \left[ 
			2 \left(x^2 - x_0^2-(y - y_0)^2 + z^2\right)
		\right]^2 
		+ 16 x^2\left( y - y_0\right) ^2 
		+ 16 z^2 \left( y - y_0\right)^2
	\right\rbrace^{-1/2}
$$
is the factor normalizing the vector \eqref{tangent_vect} to unity.

\end{document}